\documentclass[3p, 12pt, onecolumn]{elsarticle}
\usepackage[T1]{fontenc}
\usepackage{amssymb}
\usepackage{caption}
\usepackage{subcaption}
\usepackage{graphicx,wrapfig,lipsum}
\usepackage{amscd,amsmath,verbatim}
\usepackage{amsfonts,epsfig}
\usepackage{mathptmx}
\usepackage{setspace}
\usepackage{epsfig}
\usepackage{url}
\usepackage{color}

\pdfoutput=1
\begin{document}

\journal{Physics Letters A}

\begin{frontmatter}

\title{Numerical exploration of the Aging effects in spin systems}

\author{Roberto da Silva $^{1}$, T\^{a}nia Tom\'{e} $^{2}$, M\'{a}rio J. de Oliveira $^{2}$} 

\address{1 - Instituto de F\'{i}sica, Universidade Federal do Rio Grande do Sul,
Porto Alegre, Rio Grande do Sul, Brazil\\
2 - Instituto de F\'{i}sica, Universidade de S\~{a}o Paulo, S\~{a}o Paulo, S\~{a}o Paulo, Brazil
}

\begin{abstract}
An interesting concept that has been underexplored in the context of
time-dependent simulations is the correlation of total magnetization, $C(t)$%
. One of its main advantages over directly studying magnetization is that we
do not need to meticulously prepare initial magnetizations. This is because
the evolutions are computed from initial states with spins that are
independent and completely random. In this paper, we take an important step
in demonstrating that even for time evolutions from other initial
conditions, $C(t_{0},t)$, a suitable scaling can be performed to obtain
universal power laws at $T=T_{c}$. We specifically consider the significant role played
by the second moment of magnetization. Additionally, we complement the study
by conducting a recent investigation of random matrices, which are applied
to determine the critical properties of the system. Our results show that
the aging in the time series of magnetization influences the spectral
properties of matrices and their ability to determine the critical
temperature of systems.
\end{abstract}

\end{frontmatter}

\section{Introduction}

\label{Section:Introduction}

Which temporal phase of the spin system evolution contains information
regarding the criticality of a physical system? Furthermore, is it feasible
to retrieve certain initial behaviors of such a system following a period of
aging?

Particularly, in the context of time-dependent Monte Carlo (MC) simulations,
we are asking whether it is possible to observe the power law-behavior of
non-equilibrium critical dynamics \cite{JansenShort-time}, even for
short-ranged initial correlations $\left\langle \sigma _{i}\sigma
_{j}\right\rangle $\ $\neq 0$.

Let us consider the question from an even more specific point of view. Let
us suppose the Ising model on a $d$-dimensional lattice under an initial
condition where the spins are randomly and equiprobabilistically
distributed, such that $\left\langle \sigma _{i}\sigma _{j}\right\rangle =0$
and $\left\langle \sigma _{i}\right\rangle =0$. After a certain time $t_{0}$%
, we observe that $\left\langle \sigma _{i}\sigma _{j}\right\rangle \neq 0$,
but $\left\langle \sigma _{i}\right\rangle =0$ still holds true. Therefore,
if we initiate the simulations with this new initial condition, can we
obtain the same temporal power laws with the same exponents? In other words,
is aging an important factor?

An interesting measure in the context of nonequilibrium time-dependent Monte
Carlo simulations (TDMCS) is the autocorrelation (spin-spin) \cite{Huse}.
Let us consider the calculation for an arbitrary $t_{0}$: 
\begin{equation}
A(t,t_{0})=\frac{1}{N}\left\langle \sum\limits_{i=1}^{N}\sigma _{i}(t)\sigma
_{i}(t_{0})\right\rangle \text{,}
\end{equation}%
with average taken different time evolutions from different random initial
configurations.

In a highly informative and comprehensive reference by Henkel and Pleimling 
\cite{Pleimling}, it has been demonstrated that when a system is prepared at
a high-temperature and suddenly quenched to a critical temperature, the
evolution of $A(t,t_{0})$ in different spin systems suggests the presence of
a dynamical scaling behavior underlying the aging process. The same thing
can be observed in \cite{Hase2010} and recently aging phenomena in a complex
version of the two-dimensional Ginzburg-Landau equation have been observed
using the difference finite method \cite{Tauber2020}.

This behavior can be described by the following equation:

\begin{equation}
A(t,t_{0})=t_{0}^{-b}f(\frac{t}{t_{0}})  \label{Eq:Scaling_autocorrelation}
\end{equation}

Here, the parameter $b$ is defined as $b=(d-2+\eta )/z$, where d represents
the dimensionality of the system and $\eta $ is a critical exponent where $z$%
\ is the dynamic exponent, and the function $f(x)$\ exhibits the property $%
f(x)\sim x^{-\lambda _{C}/z}$\ as $x$\ approaches infinity. In this context, 
$\lambda _{C}$\ denotes the autocorrelation exponent.

An alternative approach to investigate the early stages of time evolution in
spin systems is to examine the correlation of the total magnetization. This
correlation is defined as:

\begin{equation}
C(t)=\frac{1}{N^{2}}\left\langle
\sum\limits_{i=1}^{N}\sum\limits_{j=1}^{N}\sigma _{i}(t)\sigma
_{j}(0)\right\rangle =\left\langle m(t)m(0)\right\rangle
\label{Eq:Correlation_Mario_Tania}
\end{equation}%
Here, $N$ represents the number of spins in the system, and $\sigma _{i}(t)$
denotes the spin value of spin $i$ at time $t$. The angular brackets $%
\left\langle \cdot \right\rangle $ denote the average over different time
evolutions and initial configurations. This correlation provides insights
into the relationship between the magnetization at time t and the initial
magnetization at time $0$.

Tome and Oliveira \cite{TomeOliveira1998} proposed and demonstrated that the
correlation $C(t)$\ follows a power-law behavior,\textbf{\ }%
\begin{equation}
C(t)\sim t^{\theta }\text{,}  \label{Eq:Power_law_Correlation}
\end{equation}%
when the initial magnetization $\left\langle m_{0}\right\rangle $ is zero
and the spins at time $0$ are equally likely to be $+1$ or $-1$ $p(\sigma
_{j}(0)=+1)=$ $p(\sigma _{j}(0)=-1)=\frac{1}{2}$, for $j=1,...,N$. The
exponent $\theta $ is the same as the magnetization exponent obtained in
time-dependent simulations within the context of short-time dynamics \cite%
{Zheng,Albano}. However, in those simulations, the initial conditions
require a fixed initial magnetization $m_{0}\ll 1$, which necessitates
preparation and extrapolation as $m_{0}$ approaches 0. This approach is
computationally more demanding.

At this juncture, it becomes intriguing to investigate the behavior of spin
systems when we examine the total correlation between time $t_{0}$ and a
subsequent time $t$, denoted as $C(t,t_{0})=\left\langle
m(t)m(t_{0})\right\rangle $. The correlation depends on two key factors: the
waiting time, denoted as $t_{0}$, and the observation time, indicated by $t$%
. Furthermore, we can investigate how aging impacts the determination of
criticality in the system. In this manuscript, we conveniently define the
time difference between observation and waiting time as $\Delta t=t-t_{0}$.
Aging effects become prominent when both $t_{0}\gg 1$\ and $\Delta t\gg 1$.

For this analysis, we employ a recent technique that involves constructing
Wishart-like matrices using the time evolutions of magnetization. The
spectral properties of these matrices are highly valuable in capturing the
critical properties at the initial stages of the evolution, as demonstrated
in our previous works. Therefore, we conducted computational experiments to
investigate the behavior of this method when we vary $t_{0}$ while keeping $%
\Delta t$ fixed.

In the following section, we provide comprehensive details regarding our
scaling approach for $C(t,t_{0})$, the fundamental properties of the
Wishart-like spectra, as well as pedagogical studies to substantiate the
forthcoming results in this work. Subsequently, we present our findings,
followed by concluding remarks in the final section.

\section{Methods and prepatory studies}

\label{Section:Methods}

The total correlation, as defined by Equation (\ref%
{Eq:Correlation_Mario_Tania}), assumes averages over random initial
configurations of a system with spins $\sigma _{j}(0)$, where $j=1,...,N$,
independently chosen according to: $p(\sigma _{j}(0)=+1)=$ $p(\sigma
_{j}(0)=-1)=\frac{1}{2}$ (high-temperature). In this case, if $N_{+}(t)$
represents the number of spins up and $N_{-}(t)$ represents the number of
spins down, we can express it as follows:

\begin{equation}
m(0)=m_{0}=\frac{1}{N}\left[ N_{+}(0)-N_{-}(0)\right]
\end{equation}%
$\left\langle m_{0}\right\rangle =0.$ But, $\left\langle \left[ N_{+}-N_{-}%
\right] ^{2}\right\rangle =\left\langle N_{+}{}^{2}\right\rangle
+\left\langle N_{-}{}^{2}\right\rangle -2\left\langle
N_{+}(N-N_{+})\right\rangle $. If $\left\langle N_{+}{}^{2}\right\rangle
=\left\langle N_{-}{}^{2}\right\rangle =\frac{N}{4}+\frac{N^{2}}{4}$ and $%
\left\langle N_{+}(N-N_{+})\right\rangle =N\left\langle N_{+}\right\rangle
-\left\langle N_{+}^{2}\right\rangle =\frac{N^{2}}{4}-\frac{N}{4}$.
Therefore, we have $\left\langle m_{0}^{2}\right\rangle =\frac{1}{N}$, which
implies a standard normal distribution for the initial magnetization when $%
N\gg 1$ given by:

\begin{equation}
p(m_{0})=\sqrt{\frac{N}{2\pi }}e^{-\frac{N}{2}m_{0}^{2}}
\end{equation}

However, when we consider the time evolution of different time-series
starting from these prepared initial conditions, using, for example, the
Metropolis dynamics as a prescription for these evolutions, the initial
distribution of magnetization degrades.

This degradation can be described for arbitrary $t$ by distribution: 
\begin{equation}
P(m(t))=\frac{1}{\sqrt{2\pi At^{\xi }}}\exp \left[ -\frac{m(t)^{2}}{2At^{\xi
}}\right]  \label{Eq:Complete_gaussian}
\end{equation}%
given that: 
\begin{equation}
\begin{array}{lll}
\left\langle m(t)^{2}\right\rangle -\left\langle m(t)\right\rangle ^{2} & 
\approx & \left\langle m(t)^{2}\right\rangle \\ 
&  &  \\ 
& = & A\ t^{\xi }%
\end{array}
\label{Eq:variance}
\end{equation}%
This is expected since according to short-time theory, $\left\langle
m(t)\right\rangle =0$, and for $m_{0}\approx 0$, one would expect that $%
\left\langle m^{2}\right\rangle \sim t^{\xi }$, where $\xi =\frac{(d-\frac{%
2\beta }{\nu })}{z}=\frac{2-\eta }{z}$. Janke et al. \cite{Janke} showed
that even for quenches below $T_{C}$, the relation $\left\langle
m^{2}\right\rangle \sim t^{d/z}$ holds true. Remember, for quenches below $%
T_{C}$ the ratio of the critical exponents $\beta /\nu =0$, thus the $\xi
=(d-\beta /\nu )/z$ reduces to $\xi =d/z$. This holds true for both the
nearest neighbor and long-range Ising model.

\begin{figure*}[tbp]
\begin{center}
\includegraphics[width=1.0\columnwidth]{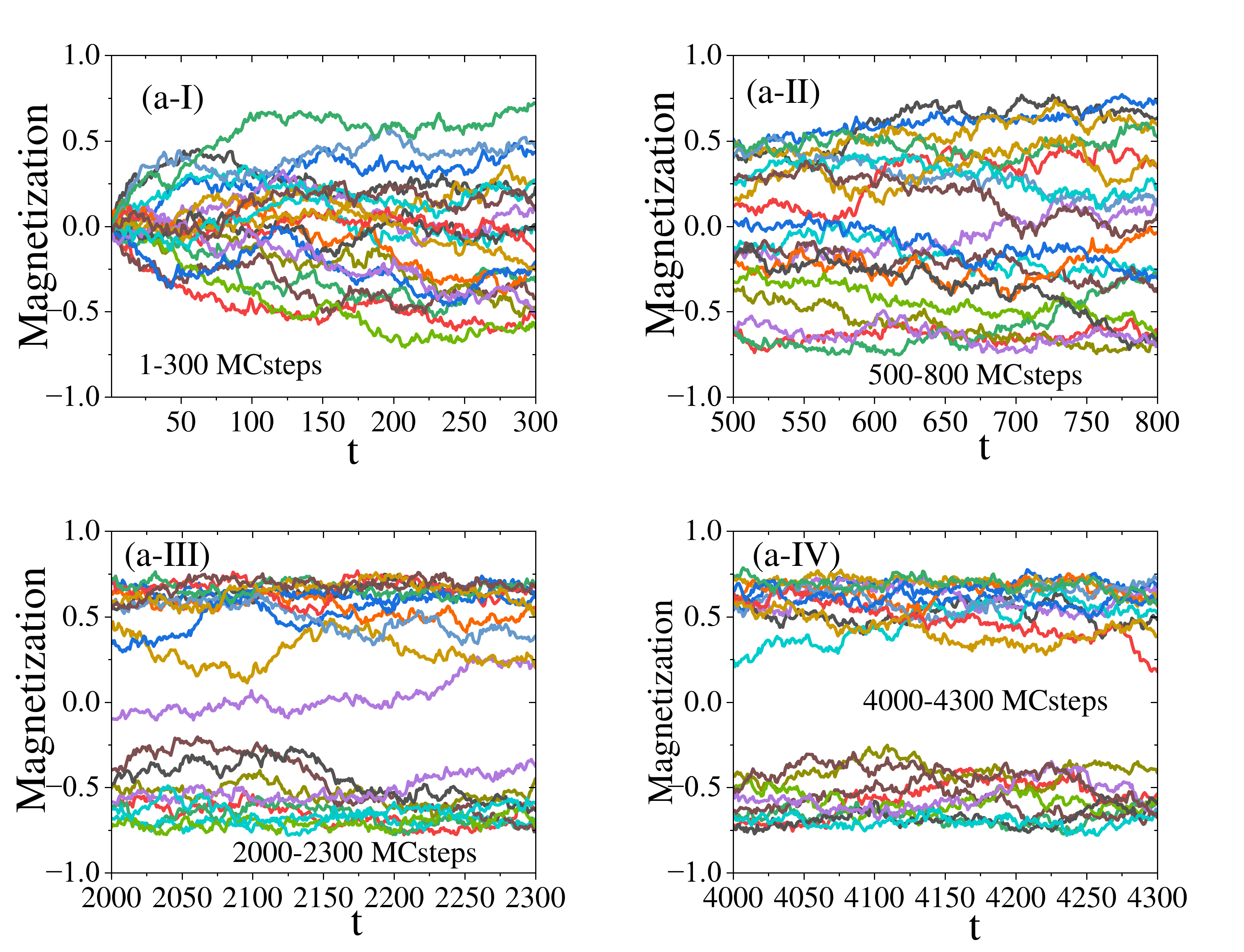} %
\includegraphics[width=0.8%
\columnwidth]{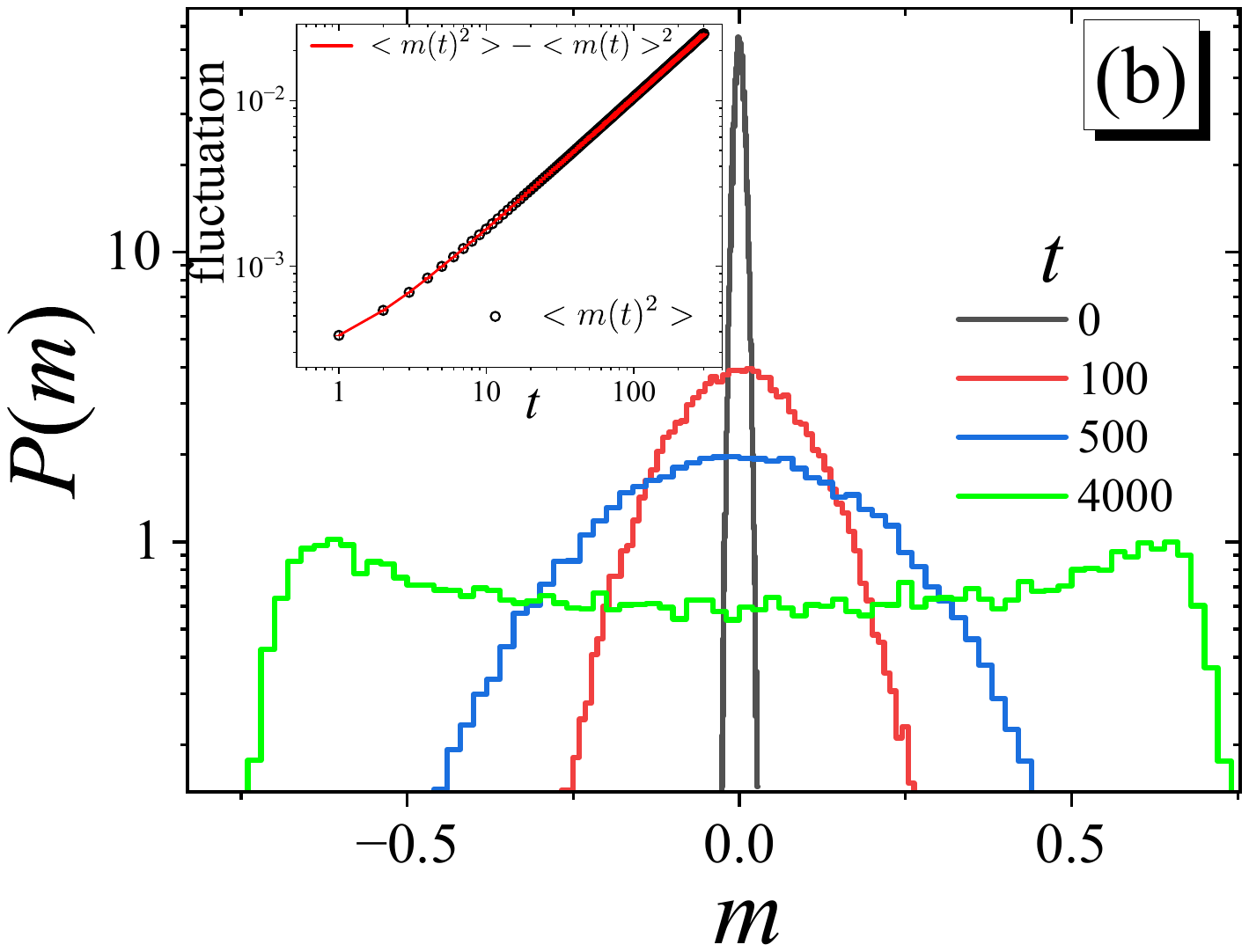}
\end{center}
\caption{(a) Aging in the time evolution of magnetization in the
two-dimensional Ising model with Metropolis dynamics for $L=100$, with time
intervals $\Delta t=300$, and initial times $t_{0}=0$, $500,2000$ and $4000$%
. (b) Histograms of magnetization for different values of $t_{0}$, following
a Gaussian distribution with variance defined by Equation \protect\ref%
{Eq:variance}. The Gaussian behavior is disrupted in the equilibrium state ($%
t_{0}\sim 4000$) when we observed that the system undergoes a slight
transition into the ordered phase due to finite size scaling. The inset plot
demonstrates that the difference $\left\langle m^{2}\right\rangle
-\left\langle m\right\rangle ^{2}$ exhibits the same power-law behavior as $%
\left\langle m^{2}\right\rangle $ in the short time regime, as $\left\langle
m\right\rangle $ is approximately 0.}
\label{Fig:Aging}
\end{figure*}

In Equation (\ref{Eq:variance}), the constant $A$ is adjustable through
fitting. Figure \ref{Fig:Aging} pedagogically illustrates this aging
phenomenon.

First, in Fig. \ref{Fig:Aging} (a), we observe different evolutions of
magnetization in the two-dimensional Ising model for various values of $%
t_{0} $, while keeping the observation time $\Delta t$ constant at 300.

We can observe histograms of magnetization for different values of $t_{0}$
in Fig. \ref{Fig:Aging} (b), following a Gaussian distribution (Eq. \ref%
{Eq:Complete_gaussian}) with variance defined by Eq. (\ref{Eq:variance}).

The Gaussian behavior is disrupted at equilibrium ($t_{0}\sim 4000$). The
inset plot in the same figure demonstrates that $\left\langle
m^{2}\right\rangle -\left\langle m\right\rangle ^{2}$ and $\left\langle
m^{2}\right\rangle $ exhibit the same power-law behavior, as $\left\langle
m\right\rangle \approx 0$. Fitting Eq. (\ref{Eq:variance}) yields the
well-known result from the literature: $\xi =0.801(1)$ for $\left\langle
m^{2}\right\rangle $, which is in complete agreement with the expected value
of $\xi =\frac{d}{z}-2\frac{\beta }{\nu z}\approx 0.802$, utilizing $%
z\approx 2.165$\ from \cite{Nightingale,Ito}, and $\frac{\beta }{\nu z}%
=0.0606$\ from \cite{Zheng}. This agreement holds true even without starting
from initial configurations with $m_{0}=0$, as is traditionally done in
computer simulations within the context of short-time dynamics.
Additionally, we obtained $A=0.00026(3)$.

From a simulation standpoint, the idea is to interrupt the simulation while
preserving the configuration at time $t_{0}$. This configuration is then
used as the initial state to calculate the correlation $C(t,t_{0})$. The
first crucial aspect is to determine if there is a finite time scaling for $%
C(t,t_{0})$ as predicted by $A(t,t_{0})$.

In other words, for very large $t_{0}$, $C(t,t_{0})$ does not depend on $%
t_{0}$. However, according to scaling theory, for sufficiently large but not
excessively large $t_{0}$, $C(t,t_{0})$ still exhibits a dependence on $%
t_{0} $.

In this paper, we aim to address this point and propose a conjecture
regarding the aging time scaling law:

\begin{equation}
C(t,t_{0})=t_{0}^{\xi }g(\frac{t}{t_{0}}),
\label{Eq:Scaling_total_correlation}
\end{equation}%
where $\xi =\frac{2-\eta }{z}$, and $g(x)\sim x^{\theta }$. Here, $\eta =%
\frac{2\beta }{\nu }$, and based on short-time theory, the exponent $\xi $
is precisely expected in the second moment of magnetization $\left\langle
m^{2}\right\rangle \sim t^{\xi }$ when starting from random initial
conditions with $m_{0}$ exactly equal to 0.

Building upon the methodology primarily developed by Henkel and Pleimling 
\cite{Pleimling}, and bolstered by valuable suggestions from anonymous
referees of this work, we will demonstrate that this quantity serves as a
correlator in momentum space and is expected to exhibit numerical scaling
according to Eq. (\ref{Eq:Scaling_total_correlation}).

Utilizing the Ising model as a simplification, our objective is to
numerically verify such scaling. We will demonstrate that considering the
magnetization distribution from Eq. (\ref{Eq:Complete_gaussian}) to select
spins is sufficient to reproduce $C(t,t_{0})$. However, we must also scale
the time by $t_{0}$ to account for the effects of non-zero spatial
correlations $\left\langle \sigma _{i}\sigma _{j}\right\rangle $.

Another important aspect addressed in this paper is the determination of the
critical properties of the system when it is out of equilibrium.
Specifically, we investigate the role of $t_{0}$ in determining the critical
properties of the system.

To examine this, we explore the effects of $t_{0}$ on the short-time
properties of the system using a recent method based on random matrices. We
developed this method to determine criticality by analyzing spectral
quantities obtained from Wishart-like matrices constructed from the time
evolutions of magnetization. In this current manuscript we will demonstrate
that the spectra is significantly influenced when large values of $t_{0}$
are used.

In the next subsection, we will provide a brief description of this method.

\subsection{Criticality in nonequilibrium regime using Wishart-like matrices
of magnetization}

\label{Subsection:Wishart}

The signature of criticality out of equilibrium seems to be even more
prominently manifested than what can be observed when uncorrelated systems ($%
T\rightarrow \infty $) are brought to finite temperatures, particularly
around $T\approx T_{C}$.

In a recent study presented at \cite{RMT2023}, we examined the response of
spectra in random matrices constructed from time evolutions of magnetization
in earlier stages of a spin system. Our findings demonstrated the influence
of criticality out of equilibrium on the spectral properties of statistical
mechanics systems. We specifically utilized the short-range two-dimensional
Ising model as a test model, as well as long-range mean-field systems \cite%
{RMT2-2023}.

To conduct such a test, we need to construct the magnetization matrix
element $m_{ij}$, which represents the magnetization of the $j$-th time
series at the $i$-th Monte Carlo step of a system with $N$ spins. Here, i
ranges from 1 to $N_{MC}$, and $j$ ranges from $1$ to $N_{sample}$.
Therefore, the magnetization matrix $M$ has dimensions $N_{MC}\times
N_{sample}$. To analyze spectral properties, an interesting alternative is
to consider not $M$, but the square matrix of size $N_{MC}\times N_{sample}$:

\begin{equation}
G=\frac{1}{N_{MC}}M^{T}M\ ,
\end{equation}%
where $G_{ij}=\frac{1}{N_{MC}}\sum_{k=1}^{N_{MC}}m_{ki}m_{kj}$, which is
known as the Wishart matrix (see for example \cite{Wishart,Wishart2,Wishart3}%
). At this stage, instead of working with $m_{ij}$, it is more convenient to
utilize the matrix $M^{\ast }$, defining its elements with the standard
variables: $m_{ij}^{\ast }=(m_{ij}-\left\langle m_{j}\right\rangle )/\sqrt{%
\left\langle m_{j}^{2}\right\rangle -\left\langle m_{j}\right\rangle ^{2}}$,
where $\left\langle m_{j}^{k}\right\rangle =\frac{1}{N_{MC}}%
\sum_{i=1}^{N_{MC}}m_{ij}^{k}$.

Therefore, $G_{ij}^{\ast }=\frac{\left\langle m_{i}m_{j}\right\rangle
-\left\langle m_{i}\right\rangle \left\langle m_{j}\right\rangle }{\sigma
_{i}\sigma _{j}}$, where $\left\langle m_{i}m_{j}\right\rangle =\frac{1}{%
N_{MC}}\sum_{k=1}^{N_{MC}}m_{ki}m_{kj}$ and $\sigma _{i}=\sqrt{\left\langle
m_{i}^{2}\right\rangle -\left\langle m_{i}\right\rangle ^{2}}$.
Analytically, if $m_{ij}^{\ast }$ are uncorrelated random variables, in this
case, the density of eigenvalues $\sigma (\lambda )$ of the matrix $G^{\ast
} $ follows the well-known Marcenko-Pastur distribution (see for example 
\cite{Sengupta}) , which is expressed as:

\begin{equation}
\sigma (\lambda )=\left\{ 
\begin{array}{l}
\tfrac{N_{MC}}{2\pi N_{sample}}\tfrac{\sqrt{(\lambda -\lambda _{-})(\lambda
_{+}-\lambda )}}{\lambda }\ ~~\text{if\ }~~~\lambda _{-}\leq \lambda \leq
\lambda _{+} \\ 
\\ 
0\ \ \ \text{otherwise}%
\end{array}%
\right.  \label{Eq:MP}
\end{equation}%
where $\lambda _{\pm }=1+\frac{N_{sample}}{N_{MC}}\pm 2\sqrt{\frac{N_{sample}%
}{N_{MC}}}.$\newline

However, for $T\neq T_{c}$, $\sigma (\lambda )$ does not follow the equation
(\ref{Eq:MP}). The behavior of $\sigma (\lambda )$ obtained from MC time
series simulated at different temperatures suggests a strong conjecture that
the average eigenvalue $\left\langle \lambda \right\rangle
=\int\nolimits_{0}^{\infty }\lambda \sigma (\lambda )d\lambda $ reaches a
minimum at the critical temperature, while the variance $var(\lambda
)=\left\langle \lambda ^{2}\right\rangle -\left\langle \lambda \right\rangle
^{2}$ exhibits an inflection point at the same critical temperature, where $%
\left\langle \lambda ^{2}\right\rangle =\int\nolimits_{0}^{\infty }\lambda
^{2}\sigma (\lambda )d\lambda $. Alternatively, a more precise
identification can be made using the negative of the derivative of the
variance:

\begin{equation}
c=-\frac{\partial var(\lambda )}{\partial T}  \label{Eq:espectral}
\end{equation}

This behavior is also observed in the Potts model \cite{Potts}. Therefore,
the idea here is to observe if such fluctuations behave differently when we
vary the starting index $i$ from $t_{0}$ to $t=t_{0}+\Delta t$, while
keeping $\Delta t=N_{MC}$ fixed.

\section{Results}

\label{Section:Results}

Our results are structured into two distinct sections. In the first section,
we provide a comprehensive justification for the scaling of Eq. (\ref%
{Eq:Scaling_total_correlation}), employing a rigorous methodology rooted in
Fourier space studies. This approach builds upon the foundational work of
Henkel and Pleimling \cite{Pleimling}, which we have extended to accommodate
Fourier space investigations. In the second section, we present the
numerical evidence substantiating this scaling behavior.

\subsection{Correlator in Fourier space}

\label{Subsection:fourier}

The two-time correlation function can be formally defined as: 
\begin{equation}
C(t,t_{0};\vec{r})=\left\langle m(t,\vec{r})m(t_{0},\vec{0})\right\rangle
-\left\langle m(t,\vec{r})\ \right\rangle \left\langle m(t_{0},\vec{0}%
)\right\rangle
\end{equation}%
where $m(t,\vec{r})$\ is the order-parameter at time $t$\ and position $r$,
and $C(t,t_{0};\vec{r})=\left\langle m(t,\vec{r})m(t_{0},\vec{0}%
)\right\rangle $\ for fully disordered initial state $\left\langle m(t,\vec{r%
})\right\rangle =\left\langle m(t,\vec{0})\right\rangle =0$.

Given the assumption of spatial translation invariance, it follows that the
two-time temporal-spatial spin-spin correlator must adhere to the following
equation: 
\begin{equation}
C(t,t_{0};\vec{r})=\kappa ^{\phi }C(\kappa ^{z}t,\kappa ^{z}t_{0};\kappa 
\vec{r})\text{,}
\end{equation}%
where $\kappa $\ is a rescaling factor, while $\phi $\ is an exponent that
can be determined by performing the scaling operation: $\kappa =\frac{1}{%
t_{0}^{1/z}}$. In this instance: 
\begin{equation}
C(t,t_{0};\vec{r})=t_{0}^{-\phi /z}C(\frac{t}{t_{0}},1;\frac{\vec{r}}{%
t_{0}^{1/z}})  \label{Eq:fi}
\end{equation}

Thus, we write: 
\begin{equation}
C(\frac{t}{t_{0}},1;\frac{\vec{r}}{t_{0}^{1/z}})=F_{C}(\frac{t}{t_{0}};\frac{%
\vec{r}}{t_{0}^{1/z}})  \label{Eq:FC}
\end{equation}

At criticality, in the equilibrium as t approaches infinity $(t\rightarrow
\infty )$, we know that $C(t,t;\vec{r})$\ must exhibit algebraic behavior as 
\begin{equation}
C(t,t;\vec{r})\sim r^{-(d-2+\eta )}  \label{Eq:Algebraic}
\end{equation}%
where $\left\vert \vec{r}\right\vert $\ represents the magnitude of $r$. By
substituting $t=t_{0}$, we can observe that: $F_{C}(1;\frac{\vec{r}}{t^{1/z}}%
)=F(\frac{r}{t^{1/z}})$. Using Eq. (\ref{Eq:fi}), we can express $C(t,t;\vec{%
r})$\ as $t^{-\phi /z}F(\frac{r}{t^{1/z}})$. Comparing this with Eq. (\ref%
{Eq:Algebraic}) yields: $F(\frac{r}{t^{1/z}})\sim \left( \frac{r}{t^{\frac{1%
}{z}}}\right) ^{-\phi }$, where $\phi =d-2+\eta $. Returning to Eq. (\ref%
{Eq:fi}), we find: 
\begin{equation}
C(t,t_{0};\vec{r})=t_{0}^{-(d-2+\eta )/z}F_{C}(\frac{t}{t_{0}};\frac{\vec{r}%
}{t_{0}^{1/z}})
\end{equation}

Dynamical symmetry arguments, however, suggest that for $t\gg t_{0}$:%
\begin{equation}
\begin{array}{lll}
F_{C}(\frac{t}{t_{0}};\frac{\vec{r}}{t_{0}^{1/z}}) & = & \left( \frac{t}{%
t_{0}}\right) ^{-\frac{\lambda _{C}}{z}}F_{C}(1;\frac{\vec{r}}{t^{1/z}}) \\ 
&  &  \\ 
& = & \left( \frac{t}{t_{0}}\right) ^{-\frac{\lambda _{C}}{z}}\mathcal{F}(%
\frac{\vec{r}}{t^{1/z}}) \\ 
&  &  \\ 
& = & \left( \frac{t}{t_{0}}\right) ^{-\frac{\lambda _{C}}{z}}\mathcal{F}(%
\frac{t_{0}^{1/z}}{t^{1/z}}\frac{\vec{r}}{t_{0}^{1/z}})%
\end{array}%
\end{equation}%
where $\lambda _{C}$\ is an exponent similar to how $\phi $\ was considered.
Thus,

\begin{equation}
C(t,t_{0};\vec{r})=t_{0}^{-(d-2+\eta )/z}\left( \frac{t}{t_{0}}\right) ^{-%
\frac{\lambda _{C}}{z}}\mathcal{F}(\frac{t_{0}^{1/z}}{t^{1/z}}\frac{\vec{r}}{%
t_{0}^{1/z}})
\end{equation}%
one has and when $r$\ equals zero, one has 
\begin{equation}
A(t,t_{0})=C(t,t_{0};0)=t_{0}^{-(d-2+\eta )/z}\mathcal{F}(0)\left( \frac{t}{%
t_{0}}\right) ^{-\frac{\lambda _{C}}{z}}\sim \left( \frac{t}{t_{0}}\right)
^{-\frac{\lambda _{C}}{z}}\text{,}
\end{equation}%
which leads to Eq. (\ref{Eq:Scaling_autocorrelation}) in the limit of large
times if $F(0)$\ is a constant. It should also be noted that we introduced $%
\lambda _{C}$, the symbol for the autocorrelation exponent, with a purpose,
and its value is determined by 
\begin{equation}
\lambda _{C}=d-z\theta \text{,}  \label{Eq:Relation_Jansen}
\end{equation}%
according to results obtained by Jansen, Schaub, and Schmittmann \cite%
{JansenShort-time}, where $\theta =(x_{0}-\frac{\beta }{\nu })/z$\
represents the initial slip exponent. Here, $\beta $\ and $\nu $\ denote the
static exponents, and $x_{0}$\ is known as the anomalous dimension of
magnetization \cite{Zheng}. For an interesting method to determine $x_{0}$,
please refer to \cite{Potts2}.

In addition, $\theta $\ is precisely the same exponent as $C(t)$\ in Eq. (%
\ref{Eq:Power_law_Correlation}). It can take on positive values (see, for
example, \cite{TomeOliveira1998,Albano,Zheng,Silvatheta}), negative values
as observed in two-dimensional tricritical points \cite%
{Silvatheta,SilvaMetamagnet}, or even very small values as seen in the
4-state Potts model due to the presence of a marginal operator \cite{Potts2}.

By defining the spatial Fourier transform of $C(t,t_{0};\vec{r})$\ as: 
\begin{equation}
\widehat{C}(t,t_{0};\vec{p})=\int_{%
\mathbb{R}
^{d}}d^{d}\vec{r}\ e^{i\vec{p}\cdot \vec{r}}C(t,t_{0};\vec{r})
\end{equation}%
where $d^{d}\vec{r}=dx_{1}dx_{2}...dx_{d},$one has:%
\begin{equation}
\begin{array}{lll}
\widehat{C}(t,t_{0};\vec{0}) & = & \int_{%
\mathbb{R}
^{d}}d^{d}\vec{r}\ C(t,t_{0};\vec{r}) \\ 
&  &  \\ 
& = & t_{0}^{-(d-2+\eta )/z}\left( \frac{t}{t_{0}}\right) ^{-\frac{\lambda
_{C}}{z}}\int_{%
\mathbb{R}
^{d}}\mathcal{F}(\frac{\vec{r}}{t^{1/z}})d^{d}\vec{r} \\ 
&  &  \\ 
& = & t_{0}^{-(d-2+\eta )/z}\left( \frac{t}{t_{0}}\right) ^{-\frac{\lambda
_{C}}{z}}t^{d/z}\int_{%
\mathbb{R}
^{d}}\mathcal{F}(\vec{u})d^{d}\vec{u}=C\ t_{0}^{\frac{(2-\eta )}{z}}\left( 
\frac{t}{t_{0}}\right) ^{\theta }%
\end{array}%
\end{equation}%
where $C=\int_{%
\mathbb{R}
^{d}}F(\vec{u})d^{d}\vec{u}$\ is supposedly a constant. Thus, by utilizing
the relation from Eq. (\ref{Eq:Relation_Jansen}) and denoting our original $%
C(t,t_{0})$\ as $\widehat{C}(t,t_{0};\vec{0})$, one obtains:

\begin{equation}
C(t,t_{0})\sim t_{0}^{\frac{(2-\eta )}{z}}\left( \frac{t}{t_{0}}\right)
^{\theta }\text{,}
\end{equation}%
by confirming the behavior described in Eq. (\ref%
{Eq:Scaling_total_correlation}) for $t\gg t_{0}$. Pleimling and Gambassi 
\cite{Gambassi} as well as Henkel et al. \cite{Henkel-Enss} have
investigated numerical results related to aging in the Fourier space of
response functions, although they did not focus on correlation as we do in
this current study.

With these results in hand, we can now delve into numerical findings
regarding this scaling and other aging effects

\subsection{Numerical Studies}

\label{Subsection:NumericalStudies}

We performed two-dimensional Monte Carlo (MC) simulations on the Ising model
in two dimensions, precisely at the critical temperature denoted as $T=T_{C}=%
\frac{2}{\ln (1+\sqrt{2})}$, employing the Metropolis dynamics with
single-flip spins.

We vary $t_{0}$. In all numerical experiments of this study, we used $L=128$%
. By starting from random initial configurations with $\left\langle
m_{0}\right\rangle =0$, we calculated $C(t,t_{0})$ considering averages over 
$N_{run}=40000$ different runs. We explored different values of $t_{0}$. The
initial question to address is determining the optimal value of $\tau $ for
which the quantity $C(t,t_{0})$\textbf{\ }$\times $\textbf{\ }$(t-t_{0}+\tau
)$\textbf{\ }follows a power law. Is $\tau $ approximately equal to $t_{0}$?

\begin{figure*}[tbp]
\begin{center}
\includegraphics[width=0.5\columnwidth]{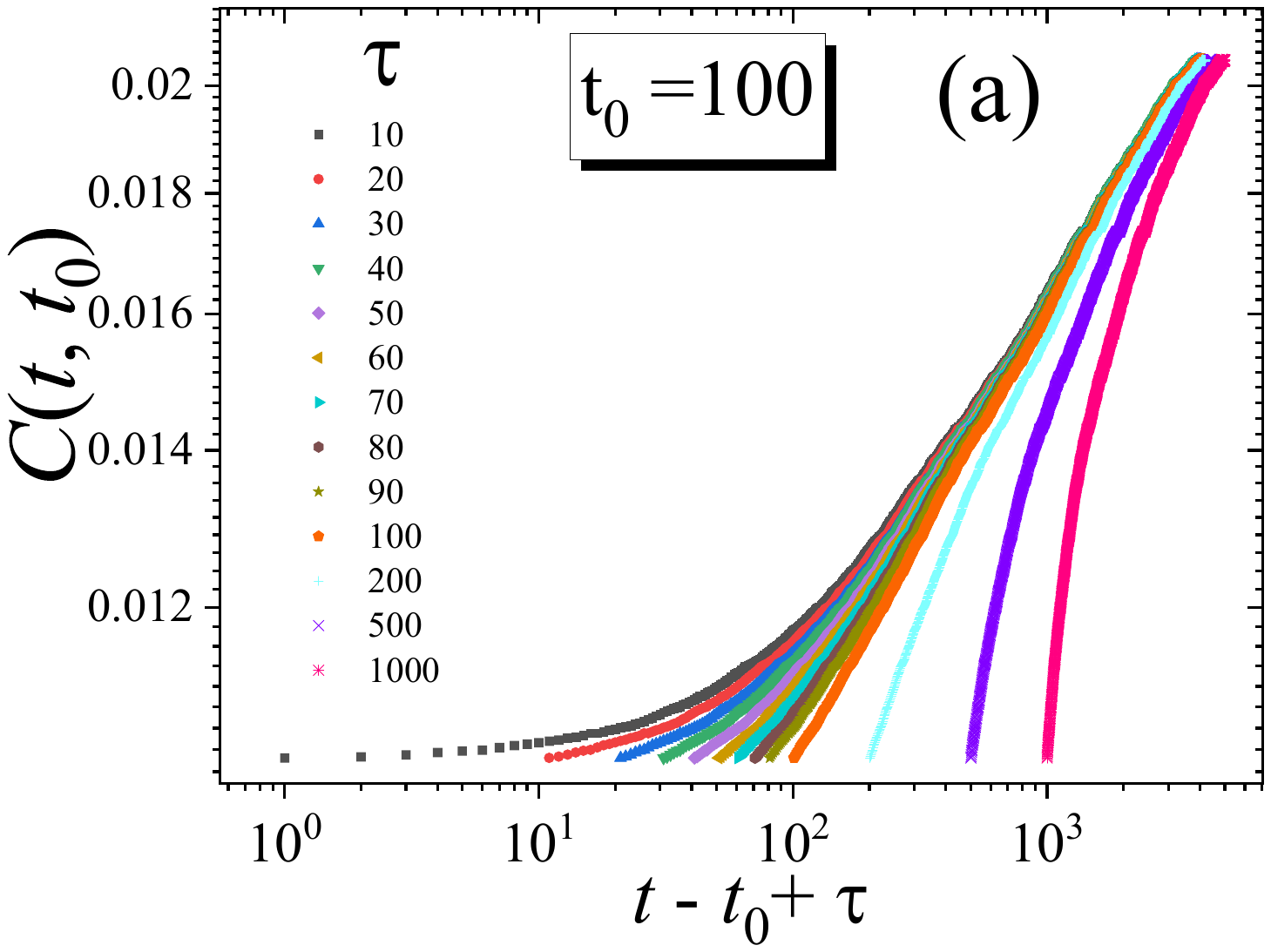}%
\includegraphics[width=0.5\columnwidth]{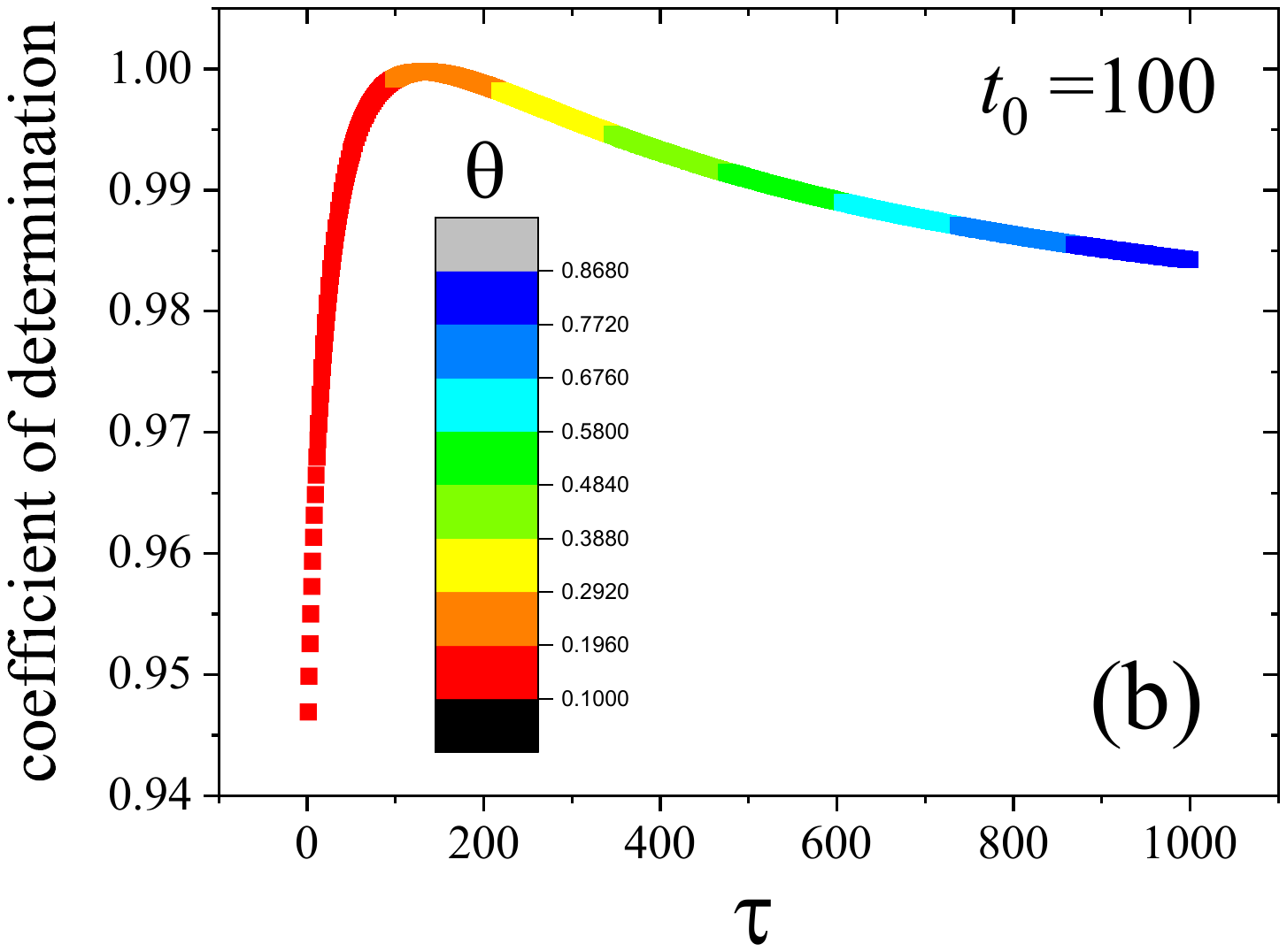} %
\includegraphics[width=0.5\columnwidth]{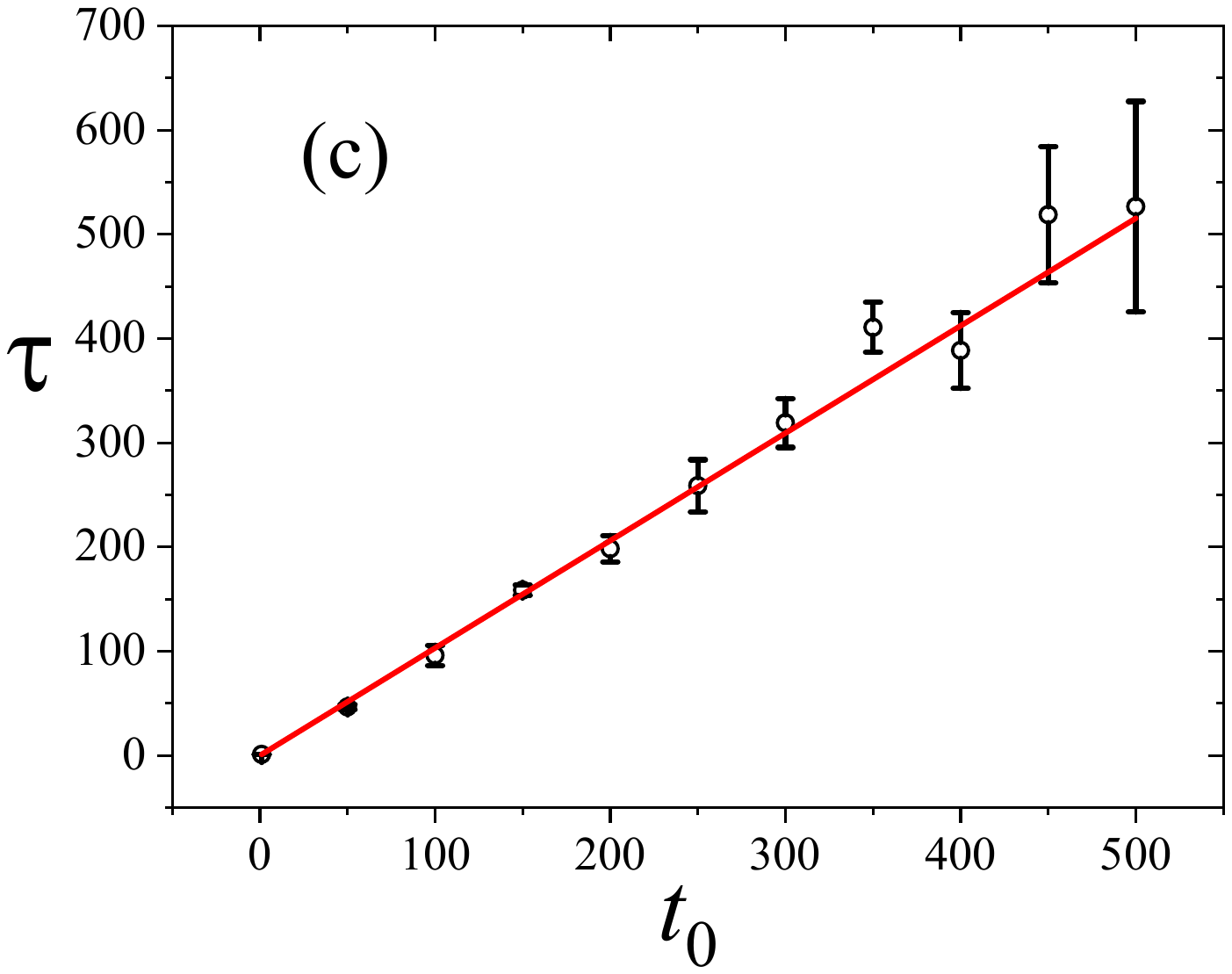}%
\includegraphics[width=0.5\columnwidth]{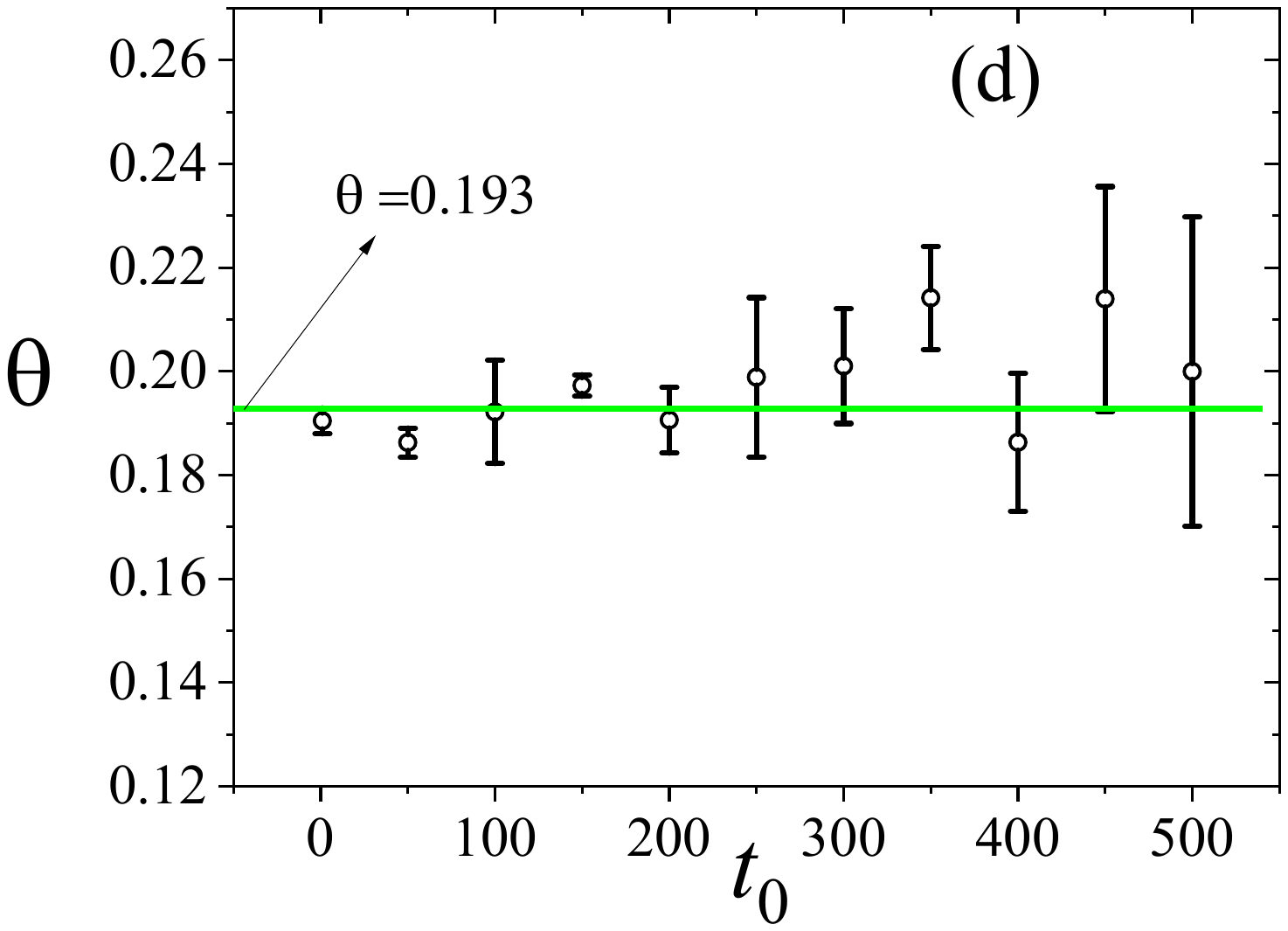}
\end{center}
\caption{(a) $C(t_{0},t)$ as a function of $t-t_{0}+\protect\tau $ for the
case $t_{0}=100$, with varying values of $\protect\tau $. We can observe
that the power law behavior occurs when $\protect\tau \approx t_{0}=100$.
(b) Coefficient of determination for the fitting with different values of $%
\protect\tau $. The maximum value occurs when $\protect\tau \approx t_{0}=100
$. The values of $\protect\theta $ are represented in different colors
according to the legend's gradient. The optimal situation (peak values)
includes $\protect\theta \approx 0.19$, as expected. (c) Optimal value of $%
\protect\tau $ as a function of $t_{0}$. (d) Corresponding value of $\protect%
\theta $ for the optimal $\protect\tau $ at each $t_{0}$.}
\label{Fig:optimization}
\end{figure*}

Thus, in order to check if $\tau $ $\approx t_{0}$, for each $t_{0}$, we
vary $\tau $ and examine the behavior of $C(t,t_{0})$ as a function of $%
t-t_{0}+\tau $ in a log-log scale for different values of $\tau $. Figure %
\ref{Fig:optimization} (a) depicts the case where $t_{0}=100$.

The power-law behavior becomes evident (qualitatively) when $\tau $\ is
approximately equal to $t_{0}$, which in this case is 100. The $\theta $\
values are visually represented using various colors to denote different
gradients, as specified in the legend. These values are derived by
conducting linear fits of $\ln C(t_{0},t)\times \ln (t-t_{0}+\tau )$\ for
each $\tau $\ as depicted in Fig. \ref{Fig:optimization} (b).

This observation is reinforced by the fact that the maximum coefficient of
determination, which approaches 1, is achieved when $\tau $\ is
approximately equal to $t_{0}$\ (100). This coefficient serves as a robust
indicator of the fitting quality, with values closer to 1 signifying a
superior fit. Its utility in the field of Statistical Mechanics has been
extensively explored for the determination of critical parameters (for
further reference, please consult \cite{Silvadetermination}).

The region of the optimal fit reveals a $\theta $\ value close to the
expected $0.19$. In Figure \ref{Fig:optimization} (c), we depict the linear
trend of the optimal $\tau $\ value, which corresponds to the highest
coefficient of determination, in relation to $t_{0}$. The linear regression
analysis produces $\tau =b$\ $t_{0}$, with $b=1.03\pm 0.02$, providing
additional evidence that $\tau $\ remains approximately equal to $t_{0}$\
regardless of the specific $t_{0}$\ value. These error bars are calculated
based on data from five different seed values.

Lastly, Figure \ref{Fig:optimization} (d) presents the corresponding $\theta 
$ values corresponding to the optimal $\tau $ values determined for various $%
t_{0}$ values.

The green line corresponds to the value observed in short-time simulations
from Ising-like models (in the same universality class) mentioned in various
references (see, for example: \cite{TomeOliveira1998,Zheng,Albano,Silvatheta}%
).

\begin{figure}[tbp]
\begin{center}
\includegraphics[width=1.0\columnwidth]{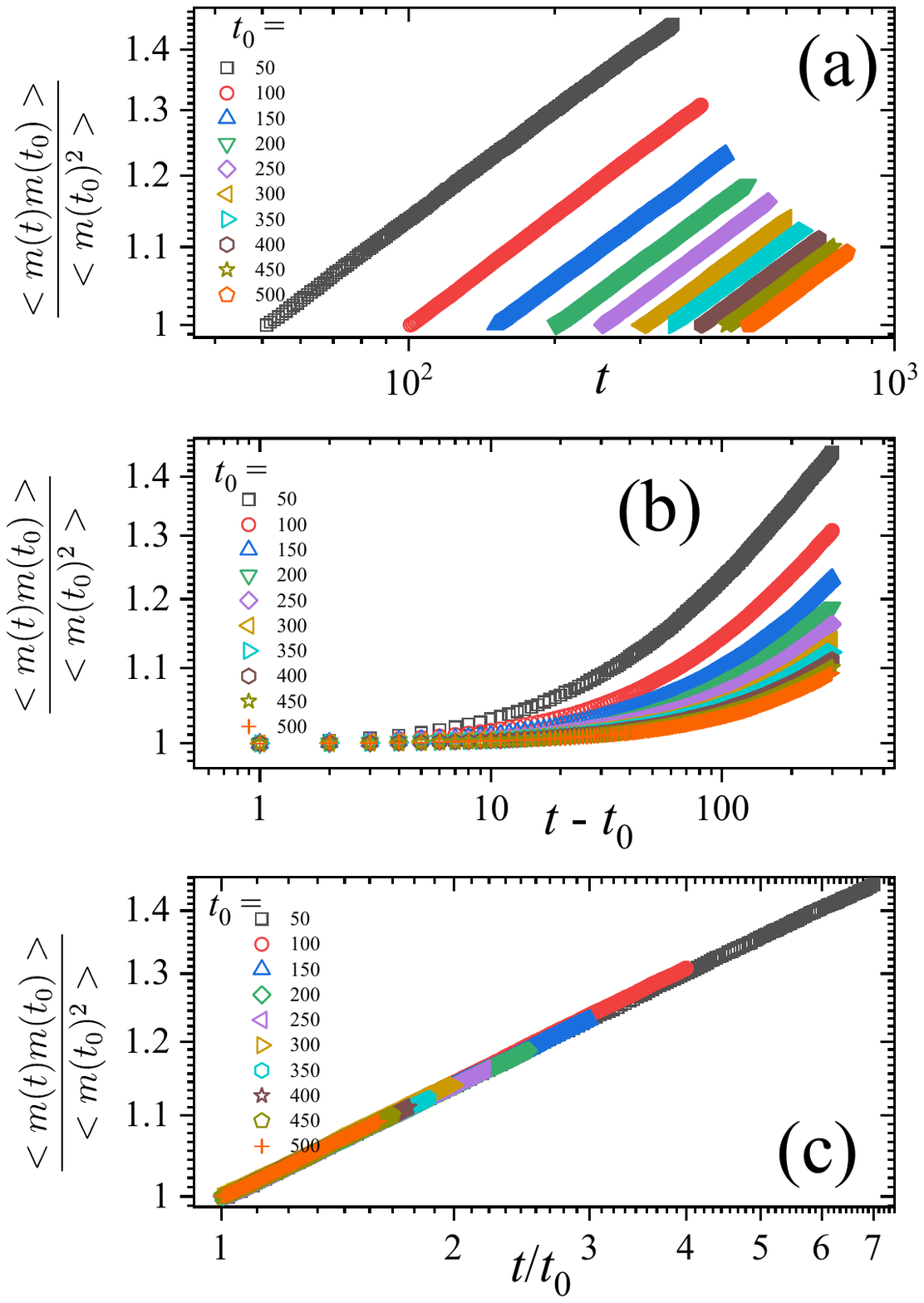}
\end{center}
\caption{Plots of the correlation divided by the initial second moment are
shown as functions of $t$ (a), $t-t_{0}$ (b), and $t/t_{0}$. An excellent
scaling is observed in plot (c), as predicted by Eq: \protect\ref%
{Eq:Scaling_total_correlation}.}
\label{Fig:Scaling}
\end{figure}

We now test the scaling relation given by Equation (\ref%
{Eq:Scaling_total_correlation}). To do so, we consider the correlation
divided by the initial second moment: 
\begin{equation}
C^{\ast }(t,t_{0})=\frac{\left\langle m(t)m(t_{0})\right\rangle }{%
\left\langle m(t_{0})^{2}\right\rangle }  \label{Eq:Correlation_normalized}
\end{equation}%
as function of $t$, $t-t_{0}$, and finally $t/t_{0}$, presented in three
different plots, all of them using log-log scale for the quantities, here
indexed by (a), (b), and (c) respectively in Fig. \ref{Fig:Scaling}. Fig. %
\ref{Fig:Scaling} (c) suggests that scaling described by Eq: (\ref%
{Eq:Scaling_total_correlation}).

These plots nicely illustrate the three defining criteria of aging \cite%
{Pleimling}: (a) slow dynamics, (b) breaking of time-translation invariance
and (c) dynamical scaling with its characteristic data collapse.

\begin{figure}[tbp]
\begin{center}
\includegraphics[width=1.0\columnwidth]{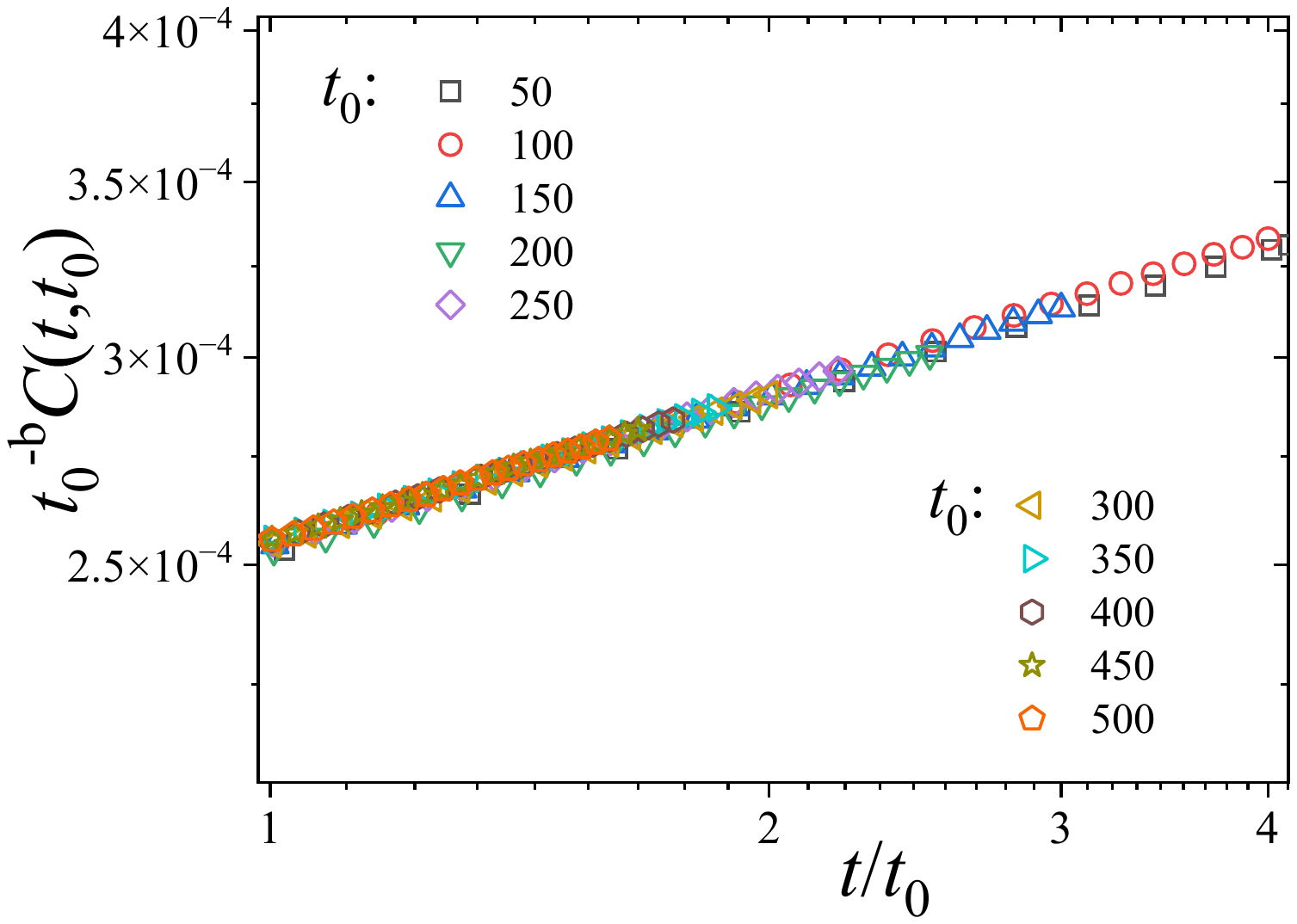}
\end{center}
\caption{Direct scaling by fitting $b$ that better corresponds to Eq. 
\protect\ref{Eq:Scaling_total_correlation}. We find $b=0.806$, which is very
similar to $\protect\xi =0.8010(4)$, expected to be the exponent of the time
evolution of $\left\langle m^{2}(t)\right\rangle $.}
\label{Fig:Direct_scaling}
\end{figure}

This scaling is performed using the quantity $\left\langle
m(t_{0})^{2}\right\rangle $\ (or $\left\langle m(t_{0})^{2}\right\rangle
-\left\langle m(t_{0})\right\rangle ^{2}$\ since $\left\langle
m(t_{0})\right\rangle =0$). Alternatively, we can perform the scaling
according to Eq. (\ref{Eq:Scaling_total_correlation}) by dividing $C(t,t_{0})
$\ by $t_{0}^{b}$\ while adjusting the value of $\xi $\ to optimize the
scaling. We also conducted a test to verify this, and the results are
presented in Fig. \ref{Fig:Direct_scaling}. We found that $b=0.806$\ is the
optimal value that matches Eq. (\ref{Eq:Scaling_total_correlation}), which
is very similar to $\xi =0.8010(4)$, the expected exponent for the time
evolution of $\left\langle m^{2}(t)\right\rangle $, by demonstrating that
the definition of $C^{\ast }(t,t_{0})$\ as outlined in Eq. (\ref%
{Eq:Correlation_normalized}) aligns seamlessly with the concepts presented
in the developments explored within subsection \ref{Subsection:fourier}.

Since the magnetization distribution at an arbitrary time $t_{0}$ follows
Eq. (\ref{Eq:Complete_gaussian}), the question is whether considering an
initial condition with magnetization distributed accordingly would yield the
same correlation $C(t,t_{0})$ as calculating it with the initial condition
that the system obtained at that time. In other words, does the obtained $%
C(t,t_{0})$ remain the same?

To explore this, we prepared systems with the initial condition described by
Eq. (\ref{Eq:Complete_gaussian}) using many different samples with different 
$m_{0}$ values, but with the condition $\left\langle m_{0}^{2}\right\rangle
=At_{0}^{\xi }$. We chose $t_{0}$ values of $50,100,200$, and $300$, which
resulted in $\sqrt{\left\langle m_{0}^{2}\right\rangle }=A^{1/2}t_{0}^{\xi
/2}=0.077$, $0.102$, $0.135$, and $\allowbreak 0.158$, respectively.

\begin{figure*}[tbp]
\begin{center}
\includegraphics[width=1.0\columnwidth]{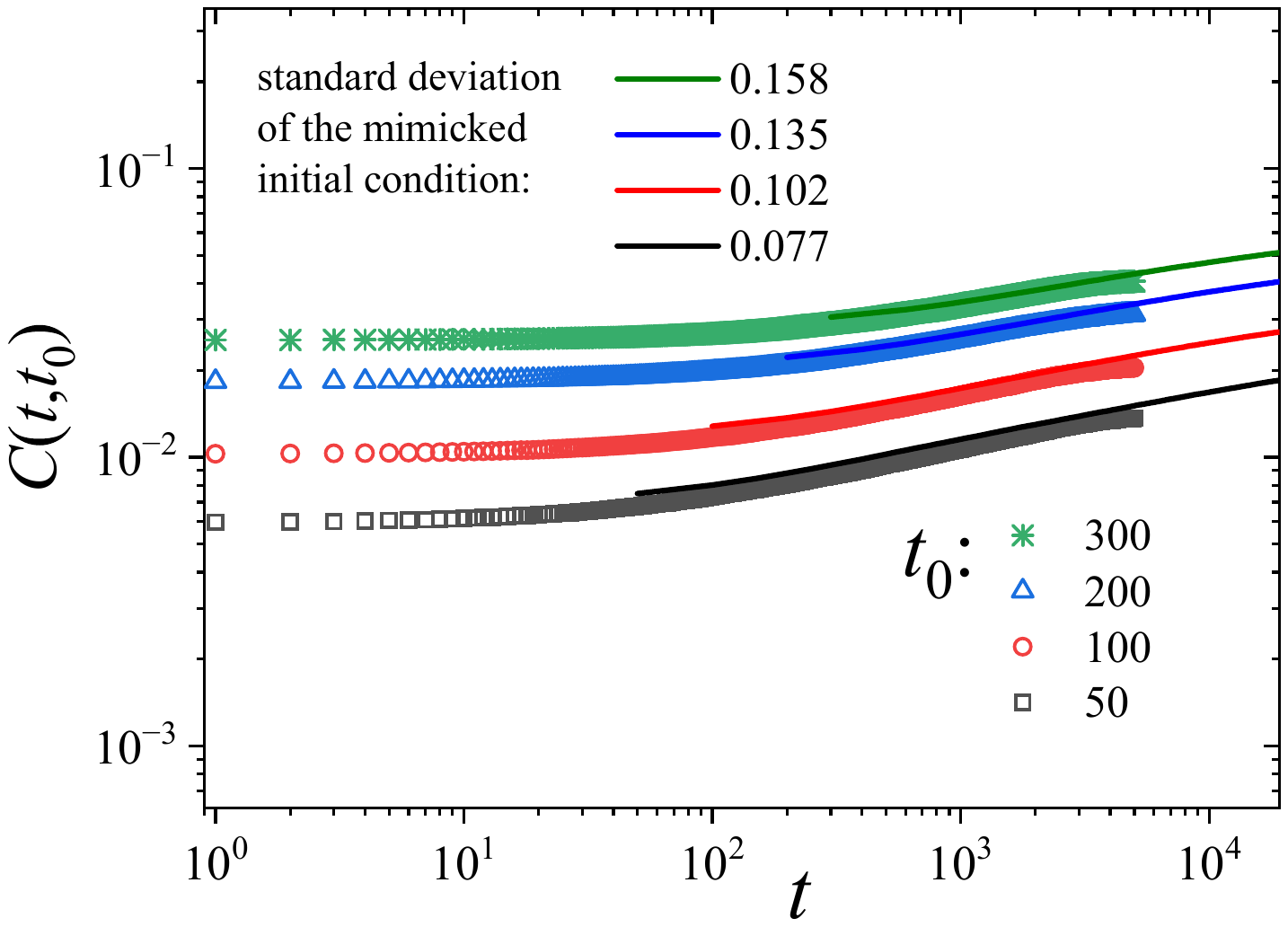}
\end{center}
\caption{Verification of $C(t,t_{0})$ for different values of $t_{0}$
(points) versus correlation by mimicking the initial distribution of spins
(continuous curves). We have scaled the time for the simulations with
mimicked initial condition by multiplying $t$ by $t_{0}$. }
\label{Fig:mimicking}
\end{figure*}

So, using these standard deviations, we generate $m_{0}$ according to Eq. (%
\ref{Eq:Complete_gaussian}), and the spins are randomly chosen with the
probability: $p(\sigma _{i})=\frac{1+m_{0}\sigma _{i}}{2}$, where $\sigma
_{i}=1,...,N$. It is important to note that $p_{-}+p_{+}=1$. We then evolve
the system and compare $C(t,t_{0})$ with the results obtained by preparing
the initial condition according to a Gaussian distribution with the
predicted variance previously established. Figure \ref{Fig:mimicking}
illustrates this comparison.

It is essential to mention that we had to scale the time by multiplying $t$
by $t_{0}$ in the case of evolutions with mimicked initial conditions. It is
interesting because it suggests that the spatial correlation of spins has an
important role, and its effects determine the time scaling of the system and
not only the distribution of magnetization to be a Gaussian according to Eq.
(\ref{Eq:Complete_gaussian}). However, we can observe that curve for $%
C(t,t_{0})$ can be reproduced if we suitably scale the time.

\subsection{Aging and random matrices}

Finally, we test the effects of aging on the spectral method sensitive to
determine the critical properties of the system. So we build matrices for $%
N_{sample}=100$, considering $\Delta t=N_{MC}=300$, and considering
different values of $t_{0}$. The result is interesting because for $t_{0}=50$%
, for example, we observe a minimum of eigenvalue mean at $T=T_{C}$ in
previous works, however when the aging is more significant, we observe a
visible deviation of such minimum as suggested by Fig. \ref%
{Fig:Average_eigenvalue} (a).

\begin{figure*}[tbp]
\begin{center}
\includegraphics[width=0.5\columnwidth]{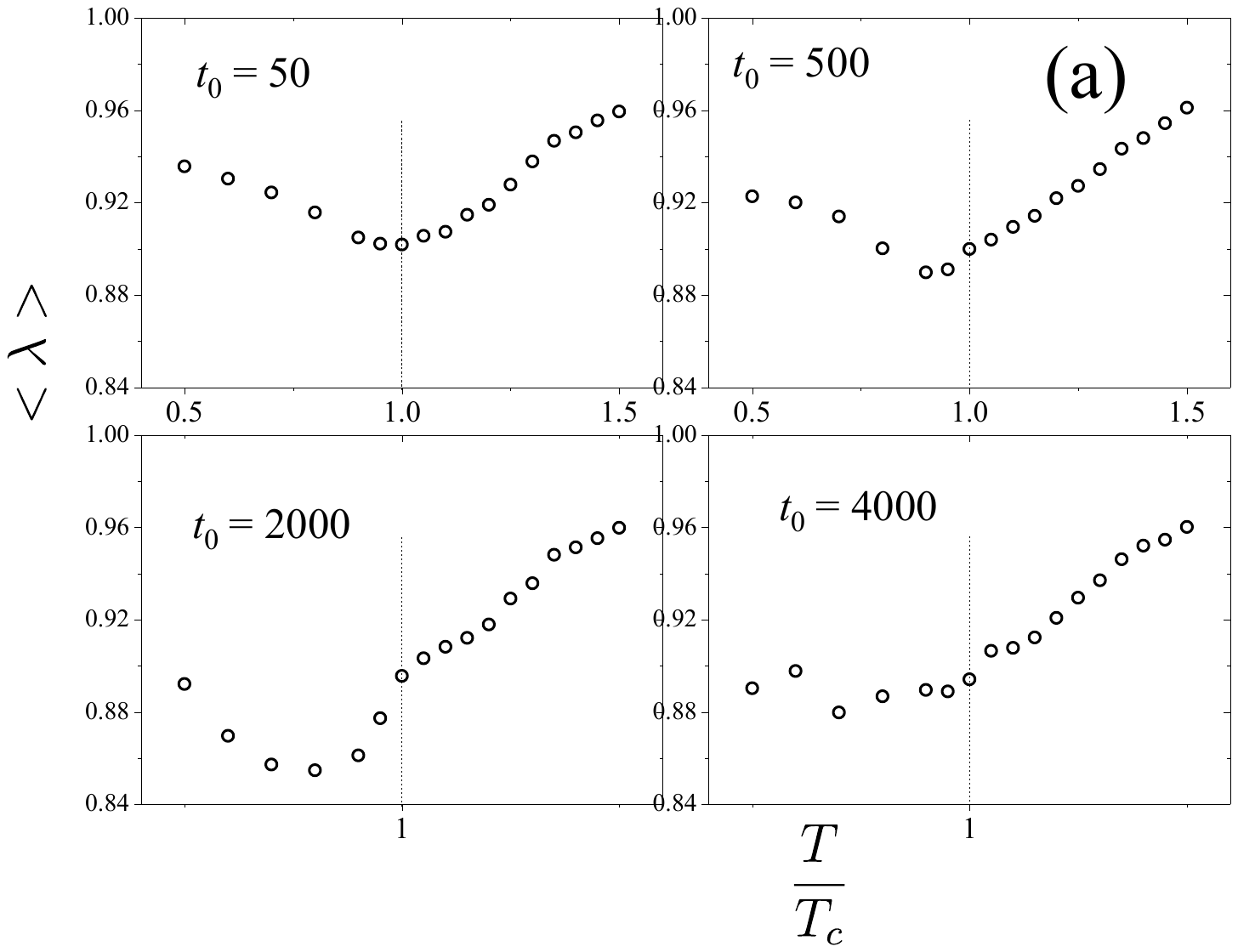}%
\includegraphics[width=0.5\columnwidth]{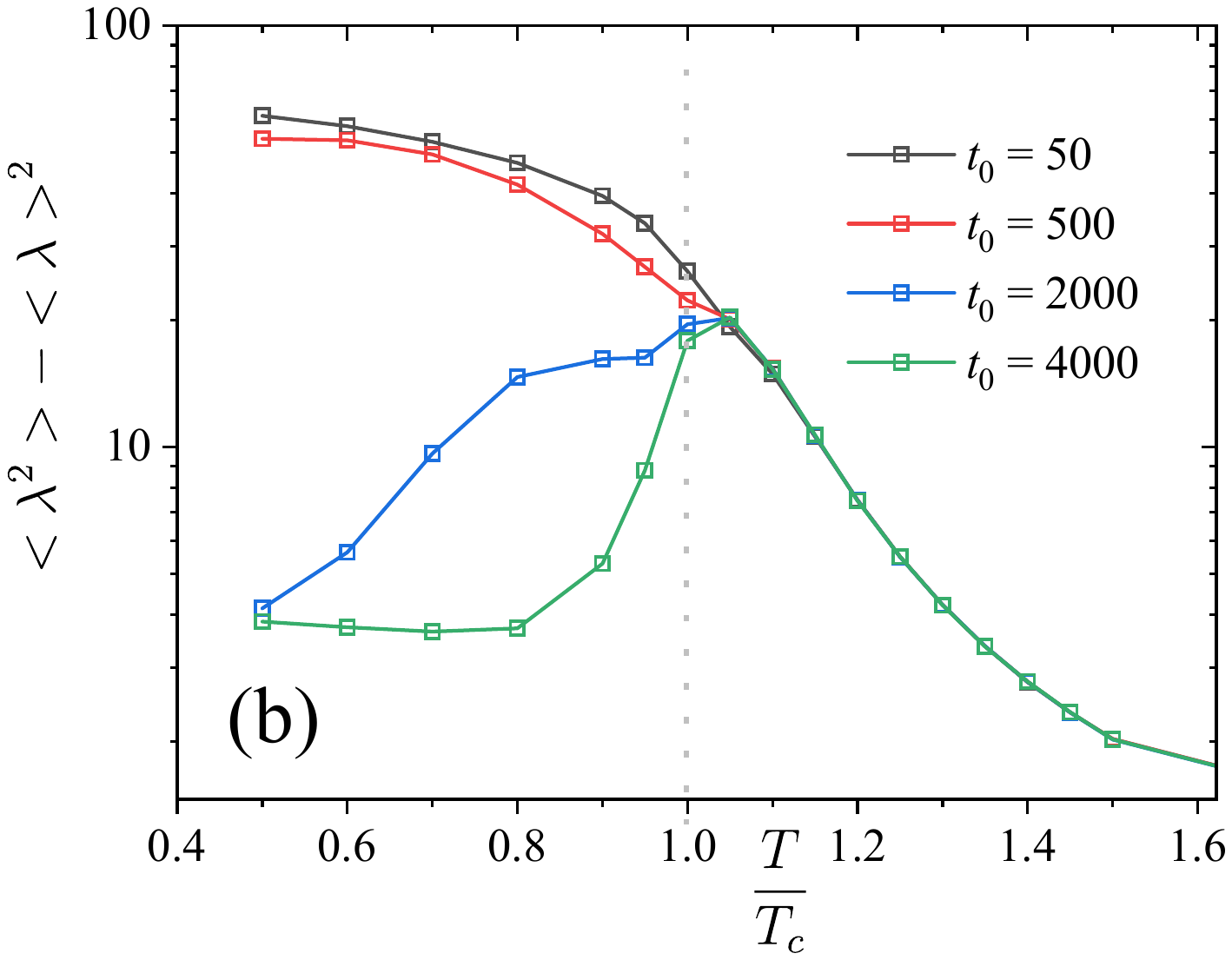} %
\includegraphics[width=0.7\columnwidth]{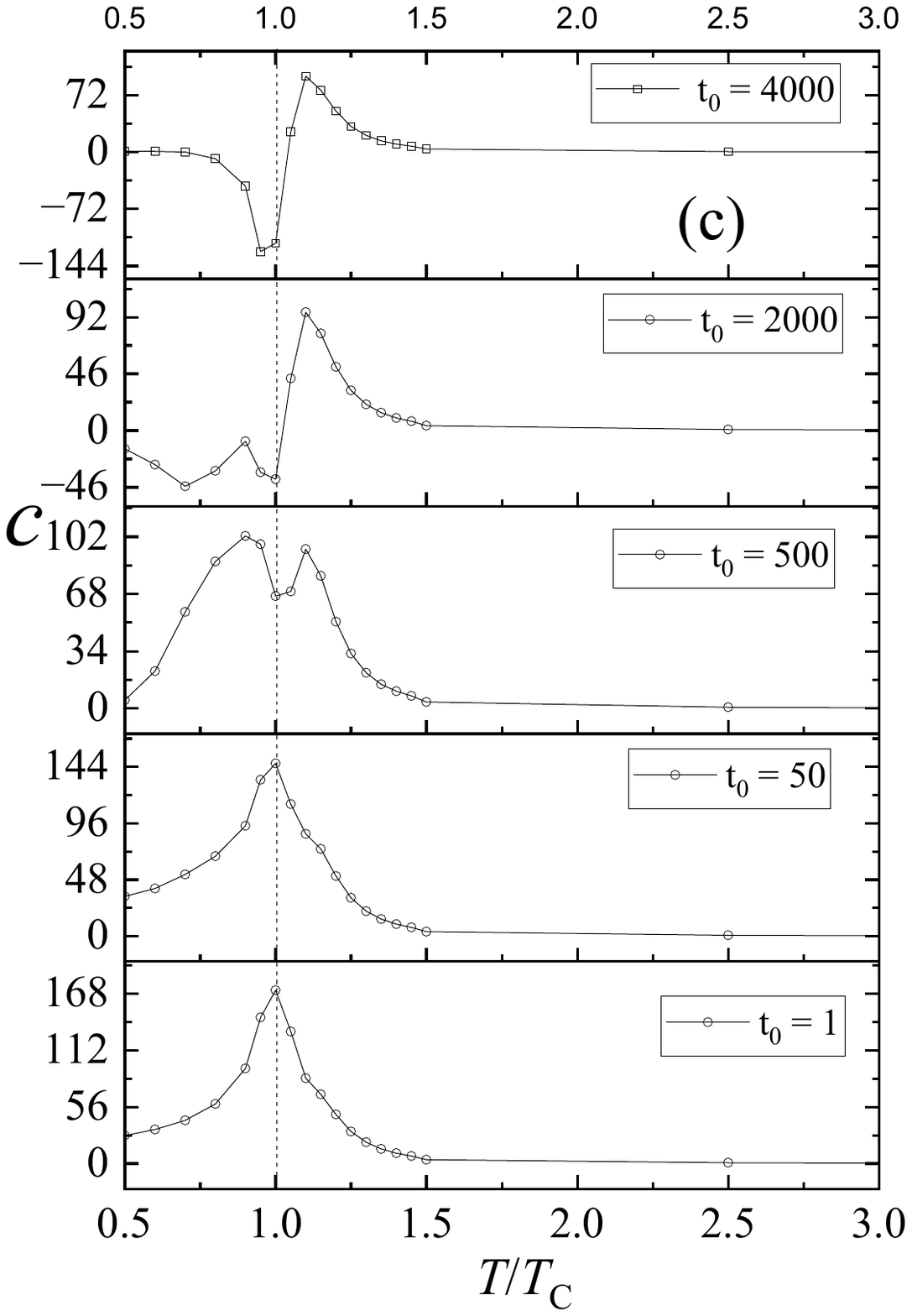}
\end{center}
\caption{(a) Eigenvalue average as a function of $T/T_{C}$. A deviation from
the minimum at $T=T_{c}$ is found for $t_{0}>50$, as can be observed. (b)
The variance of eigenvalues for different temperatures and (c) The negative
derivative of this same variance. }
\label{Fig:Average_eigenvalue}
\end{figure*}

A deviation from the minimum at $T=T_{c}$ is found for $t_{0}>50$, as can be
observed. Thus it is interesting the sensitivity of spectra considering time
series with aging. The same can be observed on the other spectral parameters
such as the variance of eigenvalues for different temperatures Fig. \ref%
{Fig:Average_eigenvalue} (b), and Fig. \ref{Fig:Average_eigenvalue} (c),
that shows the pronounced peak on the negative of the derivative of this
same variance when no aging is considered. However, after the peak ($t_{0}=1$%
), we observe a double peak ($t_{0}=500$)\ and subsequent discontinuity in
the vicinity of $T/T_{C}=1$, and there is no consensus regarding the
localization of the critical parameter. This lack of consensus is
particularly pronounced for $t_{0}>50$, where finite-size effects appear to
be significant and influence the localization.

\section{Conclusions}

We conducted a study on aging phenomena by examining the scaling behavior of
the total correlation of magnetization. Such scaling is corroborated by an
analysis of the correlator in Fourier space according to methodology
developed by Henkel and Pleimling \cite{Pleimling}. Our findings reveal an
important deviation in the scaling of the second moment of magnetization.
Moreover, we demonstrate that when considering the initial magnetization
distributed according to the Gaussian distribution expected at the time we
hypothetically started after interrupting the time-dependent simulations, we
need to scale the time appropriately to capture the correlation obtained
with this initial time.

Furthermore, we present an intriguing analysis of random matrices, which
sheds light on the expected spectra of matrices constructed from the time
evolutions of magnetization during aging. This method exhibits high
sensitivity and demonstrates how aging can impact the determination of the
critical temperature under the influence of finite size scaling effects.

Overall, our study provides valuable insights into the effects of aging on
magnetization dynamics and highlights the importance of accounting for
initial conditions and scaling considerations in such systems. It is
important to stress, that Pleimnling and Gambassi, and Henkel et. al much
before had obtained other numerical examples for studies in Fourier space
however by concerning response functions \cite{Gambassi,Henkel-Enss}.

Lastly, it is crucial to note that the irrelveant of $t_{0}$\ in the
renormalization-group sense was previously established at $T=T_{C}$\ by
Calabrese and Gambasi \cite{Calabrese}. They conducted a meticulous
comparison of correlators and responses in both direct and momentum space.
Our results align with their findings, as our short-ranged initial
correlation, in theory, should not alter the critical exponents.

It is also important to note that for temperatures below the critical point (%
$T<T_{C}$), Janke and colleagues \cite{Janke}, have established that scaling
laws, which remain invariant regardless of the selection of $t_{0}$, are
likely associated with the behavior $\left\langle m(t)^{2}\right\rangle \sim
t^{d/z}$, even in the context of long-range systems.

\section*{Acknowledgments}

This paper would not have been what it is without the invaluable
philanthropic support of the anonymous referees. Their contributions have
been so significant that they could rightfully be considered co-authors of
this manuscript. Our heartfelt gratitude goes out to them!

\section*{CRediT authorship contribution statement}

All authors conceived and designed the analysis, performed formal analysis,
wrote the paper, elaborated the algorithms, analysed the results, and
reviewed the manuscript.

\section*{Declaration of competing interest}

The authors declare that they have no known competing financial interests or
personal relationships that could have appeared to influence the work
reported in this paper

\section*{Funding}

R. da Silva would like to thank CNPq for financial support under grant
number 304575/2022-4.

\end{document}